%% file: paper.tex
\documentclass[pdflatex,sn-mathphys-num]{sn-jnl}

\usepackage{graphicx}%
\usepackage{multirow}%
\usepackage{amsmath,amssymb,amsfonts}%
\usepackage{amsthm}%
\usepackage{mathrsfs}%
\usepackage[title]{appendix}%
\usepackage{xcolor}%
\usepackage{textcomp}%
\usepackage{manyfoot}%
\usepackage{booktabs}%
\usepackage{algorithm}%
\usepackage{algorithmicx}%
\usepackage{algpseudocode}%
\usepackage{listings}%

\begin{document}

\title[Leveraging AI to Advance Science and Computing Education across Africa]{Leveraging AI to Advance Science and Computing Education across Africa: Challenges, Progress and Opportunities}

\author[1,2]{\fnm{George} \sur{Boateng}}\email{jojo@kwame.ai}

\affil[1]{\orgname{Kwame AI Inc.}, \country{U.S}}

\affil[2]{\orgname{ETH Zurich}, \country{Switzerland}}

\abstract{Across the African continent, students grapple with various educational challenges, including limited access to essential resources such as computers, internet connectivity, reliable electricity, and a shortage of qualified teachers. Despite these challenges, recent advances in AI such as BERT, and GPT-4 have demonstrated their potential for advancing education. Yet, these AI tools tend to be deployed and evaluated predominantly within the context of Western educational settings, with limited attention directed towards the unique needs and challenges faced by students in Africa. In this chapter, we discuss challenges with using AI to advance education across Africa. Then, we describe our work developing and deploying AI in Education tools in Africa for science and computing education: (1) SuaCode, an AI-powered app that enables Africans to learn to code using their smartphones, (2) AutoGrad, an automated grading, and feedback tool for graphical and interactive coding assignments, (3) a tool for code plagiarism detection that shows visual evidence of plagiarism, (4) Kwame, a bilingual AI teaching assistant for coding courses, (5) Kwame for Science, a web-based AI teaching assistant that provides instant answers to students’ science questions and (6) Brilla AI, an AI contestant for the National Science and Maths Quiz competition. Finally, we discuss potential opportunities to leverage AI to advance education across Africa.}


\maketitle

\input{content.tex}

\bmhead{Acknowledgments}
These works have been supported over the years with grants from the Processing Foundation, the African Union, the Africa Prize for Engineering Innovation program of the U.K.'s Royal Academy of Engineering, Dartmouth College, and ETH Zurich. The following generative AI tools were used in preparing this manuscript: ChatGPT, Perplexity, and Consensus.

\bibliography{refs}

\end{document}

%% file: content.tex
\section{Introduction}
In Africa, a significant portion of students grapples with formidable educational barriers arising from a multitude of challenges, including limited access to essential resources such as computers \cite{UNESCO2020Digital}, internet connectivity \cite{iea2023, UNESCO2020Digital}, reliable electricity \cite{dw2020}, and a shortage of qualified teachers \cite{UNESCO2020}. For instance, as of 2018, the average student-teacher ratio in Sub-Saharan Africa stood at 35:1, starkly contrasting with the more favorable ratio of 14:1 observed in Europe \cite{UNESCO2020}. The absence of these resources impedes the quality and accessibility of education across the continent.

Despite these challenges, recent strides in Artificial Intelligence (AI) technology, exemplified by sophisticated systems like BERT \cite{devlin2019} and GPT-4 \cite{achiam2023}, have showcased their potential to revolutionize education globally. Major players in the global EdTech landscape, such as Duolingo \cite{duolingomax}, Quizlet \cite{q-chat}, Chegg \cite{cheggmate}, and KhanAcademy \cite{khanmigo}, have actively integrated these AI advancements into their platforms to enhance learning experiences. However, the deployment and evaluation of these AI systems predominantly occur within the context of Western educational settings, with limited attention directed towards the unique needs and challenges faced by students in Africa. For example, the release of GPT-4 in March 2023 featured various academic exams as benchmarks with none from Africa \cite{achiam2023}. This oversight underscores the tendency for the African continent to be marginalized in the application of cutting-edge AI advancements, with little consideration given to the diverse educational landscapes and requirements of African students. As a consequence, the potential of AI to address the educational inequities and barriers prevalent in Africa remains largely untapped, highlighting the urgent need for greater inclusivity and tailored solutions that account for the specific needs and challenges faced by students on the continent.

In this chapter, we discuss challenges with using AI to advance education across Africa. Then, we describe our work developing and deploying AI in Education (AIED) tools in Africa for science and computing education: (1) SuaCode, an AI-powered app that enables Africans to learn to code using their smartphones, (2) AutoGrad, an automated grading, and feedback tool for graphical and interactive coding assignments, (3) a tool for code plagiarism detection that shows visual evidence of plagiarism, (4) Kwame, a bilingual AI teaching assistant for coding courses, (5) Kwame for Science, a web-based AI teaching assistant that provides instant answers to students’ science questions and (6) Brilla AI, an AI contestant for the National Science and Maths Quiz competition (NSMQ). Finally, we discuss potential opportunities to leverage AI to advance education across Africa and then conclude.

\section{Challenges} 
One of the key challenges with leveraging AI to improve education in Africa is access to resources (like computers) and infrastructure (Internet, electricity) to make use of AIED tools. Across sub-Saharan Africa, Internet data is expensive \cite{dw2020} and electricity is either not available or erratic in some countries \cite{iea2023}. According to UNESCO, 11\% of students in sub-Saharan Africa have access to household computers, and only 18\% have access to the Internet \cite{UNESCO2020Digital}. So, how can students in a rural part of Ghana, for example, access and use an AI tutor if they do not have computers or smartphones or constant electricity? If these tutors are deployed to students in urban areas who have the necessary resources and infrastructure, it could increase the divide between the haves and the have-nots, further exacerbating the existing inequities in the education sector in various African countries.

A lack of regulatory support for innovation could hamper efforts to deploy AIED tools. For example, several Senior High Schools in Ghana are boarding schools that do not allow their students to have access to their devices (smartphones and computers) \cite{boateng2023}. Since students spend about 9 months of the 12 months in school, they mostly do not have access to devices. Only a few private schools have computers in their libraries for students to use. This issue is a significant barrier to the potential impact of AIED tools on students at the senior high school level. How do you give access to an AI tutor to students if the Ghana Education Service, for example, restricts access to smart devices in boarding schools at the secondary school level?

There is heterogeneity in educational systems across different parts of Africa making it difficult to deploy AIED tools scalably. For example, Anglophone West Africa has the West African Secondary School Certificate Exam (WASSCE) and Kenya has the Kenya Certificate of Secondary Education (KCSE). Some countries use English whereas others use French as their official language of instruction. Yet, several local languages abound which may need to be supported to reach students who have low literacy in official languages. Hence, AIED tools developed in one country or region of the continent would need to be adapted to the local educational system of other countries. Considering there are 54 countries in Africa, this process would also entail jumping through different regulatory hoops making it time-consuming and costly.

Educational materials in various parts of Africa needed as input for building AIED tools tend to be available in hardcopy format which poses challenges for digitization. One challenge we encountered revolved around the formatting of scientific and mathematical symbols and equations. Regrettably, the open-source optical character recognition (OCR) technologies we experimented with proved inadequate in accurately extracting these symbols \cite{boateng2023}. Consequently, manual intervention became necessary to ensure an additional layer of quality control. There is a pressing need for further advancements in developing systems capable of seamlessly converting scanned scientific documents into outputs that maintain correct representations, facilitating compatibility with standard formats such as markdown. Until such solutions are realized, substantial time and financial investments will continue to be essential for generating high-quality usable data.

One of the critical challenges facing the development and deployment of AIED tools is the pervasive lack of representative data, particularly from regions in the Global South in pretrained AI models. There are challenges accessing copyrighted, local textbooks to build AIED tools \cite{boateng2023}. The absence of diverse and inclusive datasets undermines the effectiveness and fairness of AI technologies, leading to biased outcomes that fail to adequately address the needs and realities of communities in these regions. Some AI tools are built, evaluated, and deployed generally and also for educational use cases without incorporating data from Africa such as GPT-4’s evaluation of missing education data from Africa  \cite{achiam2023}. These result in biased systems that do not well in the African context. Examples include speech-to-text systems that do not well for African accents, or text-to-speech systems that do not speak with an African accent \cite{boateng2023nsmqai}. If such AI systems are integrated into an AI tutor for African students, for example, they will not work well, defeating the goal of the AI tutor.

Students could end up using these AIED tools as a crutch either for cheating or just being overreliant on them such as to solve homework assignments, write their essays for them and even cheat in national exams \cite{wasscecheating}. If students use these tools in this way, it defeats the pedagogical goals of the assigned homework. Relying excessively on AIED tools for completing homework assignments poses a significant risk of undermining the educational objectives set by instructors. When students resort to these tools as a crutch, they miss out on the opportunity to engage critically with the material and develop essential problem-solving skills. Instead of grappling with the complexities of the subject matter, students may opt for the path of least resistance, simply inputting their assignment prompts into AI platforms and passively accepting the generated solutions. This not only diminishes the intellectual rigor of the learning process but also fosters a culture of academic dishonesty.

Generative AI models that power AIED tools have an issue of sometimes confidently generating responses that are factually incorrect, referred to as “Hallucination” which could affect their utility in improving education. In instances where students rely heavily on AI-generated content for learning and study purposes, the propagation of misinformation can undermine the integrity of their educational experiences. Instead of fostering a deep understanding of the subject matter, the presence of inaccuracies may cultivate a false sense of comprehension, ultimately hindering students' ability to engage critically with the material.

\section{Progress: AIED Solutions}
In this section, we discuss AIED tools that we developed and deployed for science and computing education, and how they address some of the challenges previously discussed.

\subsection{SuaCode}
\begin{figure}[ht]
    \centering
    \includegraphics[width=\linewidth]{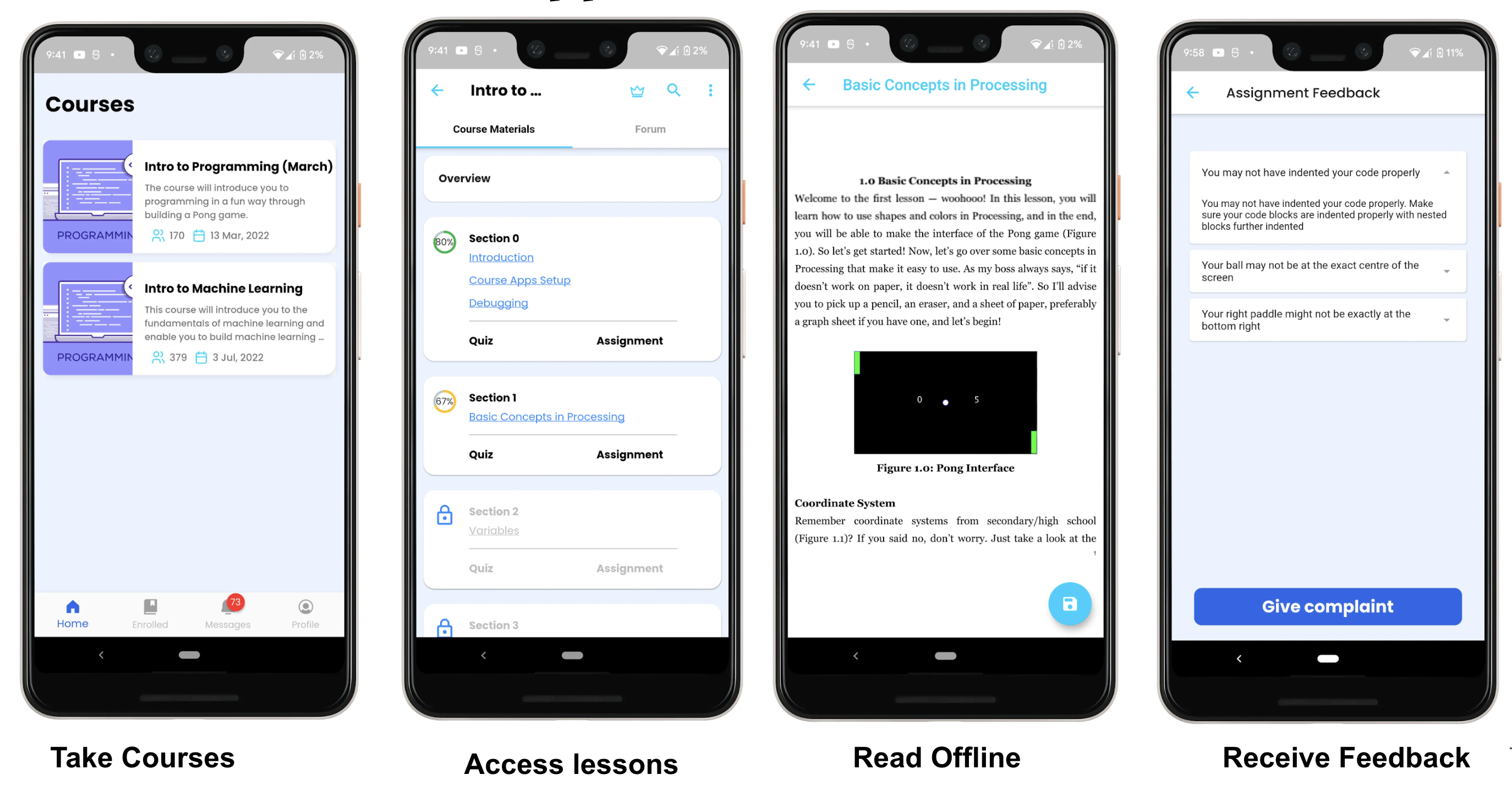}
    \caption{Screenshots of the SuaCode App with Course Materials and Assignment Feedback}
    \label{fig:suacode1}
\end{figure}

\begin{figure}[ht]
    \centering
    \includegraphics[width=\linewidth]{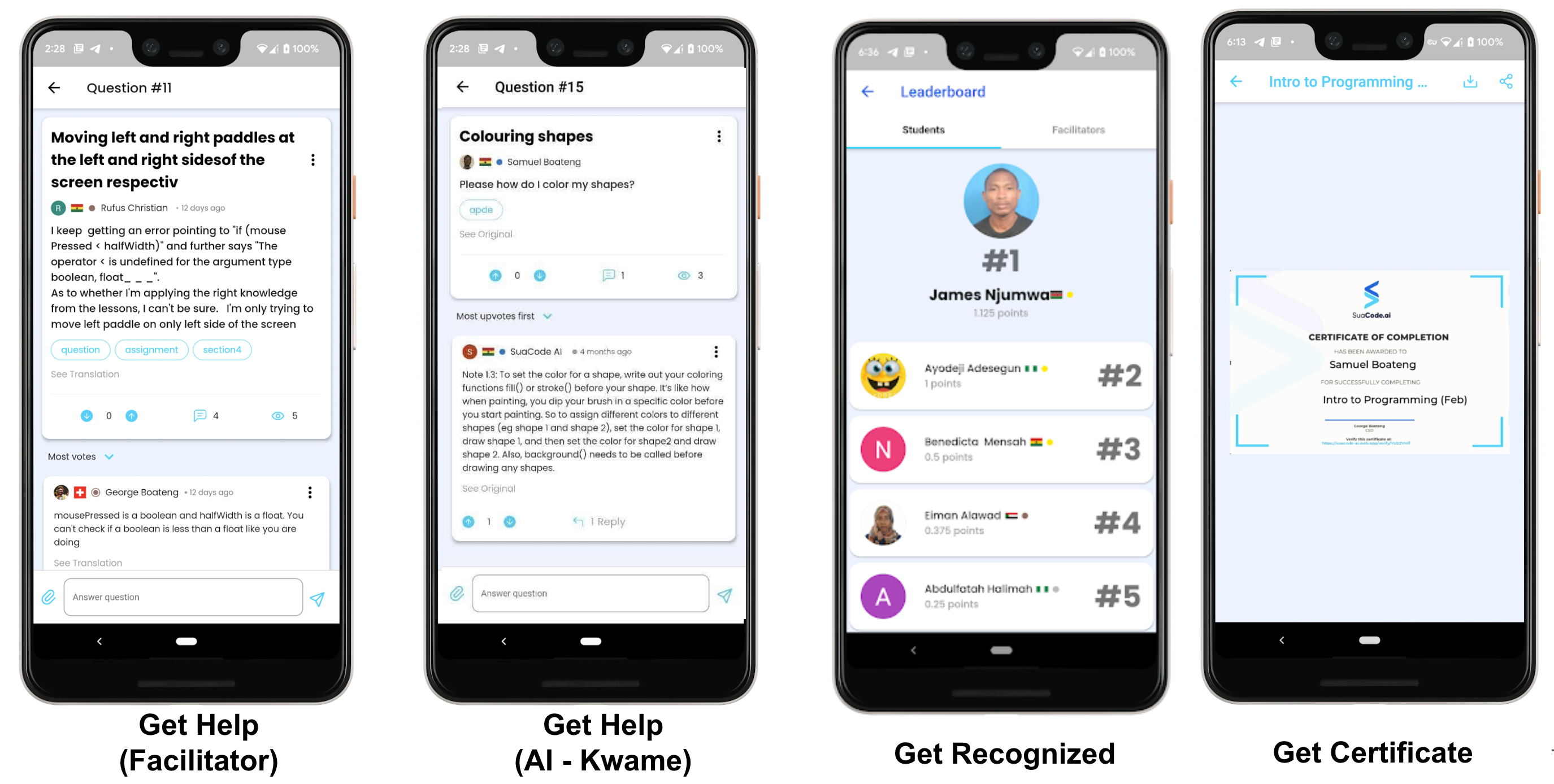}
    \caption{Screenshots of the SuaCode App with Forum, Leaderboard, and Certificate}
    \label{fig:suacode2}
\end{figure}

SuaCode \footnote{\href{http://suacode.ai/}{http://suacode.ai/}} is an AI-powered smartphone-based application that enables students across Africa to learn to code using smartphones (Figures \ref{fig:suacode1} and \ref{fig:suacode2}). The app has monthly online coding courses with lesson notes, exercises, quizzes, and fun coding assignments in English and French which cover official languages across Africa. We use simplified lesson notes designed for offline reading to reduce Internet data usage since they are expensive across Africa \cite{dw2020}. Our courses adopt a project-based learning approach where learners build and interact with a game on their phones as assignments. Each section of the course ends with multiple-choice quizzes and a coding assignment. The assignments are graded by our automated grading software, AutoGrad \cite{annor2021} which also checks for plagiarism \cite{john2021} and provides detailed individualized explanations for wrong answers with an option for learners to submit a complaint in cases where they disagree with the provided answers and feedback. Our course-specific, AI-powered, human-in-the-loop forum is designed to allow learners to ask questions  (even anonymously) and get quick and accurate answers from their peers, facilitators, and our AI teaching assistant, Kwame \cite{boateng2021b}. Kwame enables learners to get individualized learning support so they do not quit when it gets tough. Each person’s name shows by the side an earned badge to encourage helpful engagement. Each of the 3 badges (bronze, silver, and gold) is received based on a “helpfulness score” that we calculate using various metrics such as upvotes on questions and answers contributed by each learner. Our leaderboard celebrates helpful learners and facilitators thereby encouraging others to be helpful. Students who complete receive certificates and mentoring by African tech professionals in companies such as Google and Amazon enabling our learners to receive advice from experienced individuals who have similar backgrounds as them. SuaCode has over 2.5K learners across 43 African countries and 117 globally.

We created SuaCode to address the problem that less than 1\% of African children leave school with basic coding skills with several jobs struggling to fill IT-related roles despite Africa being home to the youngest workforce in the world \cite{sap2016}. In 2017, while running the 4th edition of our annual innovation bootcamp, Project iSWEST, in Ghana, we noticed from our pre-survey that out of our 27 students, 25\% had laptops, yet 100\% had smartphones. This situation of limited access to computers led us to modify our coding course and deliver it using smartphones, the first of its kind in Ghana. Our students built pong games on their phones with several students working on coding assignments while in traffic \cite{boateng2018}. Realizing the potential of our smartphone-based course, in 2018, we created SuaCode, a smartphone-based online coding program aiming to teach millions across Africa how to code by exploiting the proliferation and untapped capabilities of smartphones  \cite{boateng2019}. 

\begin{figure}[ht]
    \centering
    \includegraphics[width=0.8\linewidth]{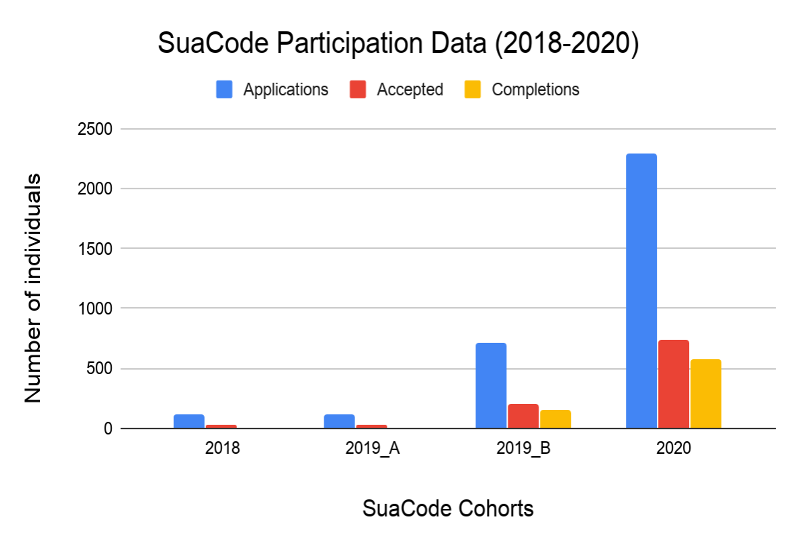}
    \caption{Growth of SuaCode between 2018 and 2020}
    \label{fig:suacode_growth}
\end{figure}

Between 2018 and 2020, we ran 4 pilots of SuaCode while growing exponentially (Figure \ref{fig:suacode_growth}) that reached 3K learners across 69 countries (42 in Africa) \cite{boateng2019, boateng2020, boateng2021}. The 2018 version focused on students in Ghana and had over 30 students enrolled with only 23\% completing \cite{boateng2019}. It was delivered using Google Classroom as the course management platform (forum and assignment submission), Google Docs for lesson notes and APDE for the coding app. We had 4 lessons and assignments, and an optional project at the end. The first version in 2019 also focused on Ghanaians. The second version in 2019 expanded beyond Ghana to all Africans and was dubbed SuaCode Africa \cite{boateng2021}. It had 709 applicants across 37 African countries. We introduced an acceptance criterion (submit the first 2 assignments) to filter for highly motivated students. Out of that, 210 were accepted and 151 completed resulting in a 72\% completion rate. The 2020 version was dubbed SuaCode Africa 2.0 with the course being offered in both English and French \cite{boateng2020, suacodeafrica2}. We used Piazza for that cohort which had a better forum experience than Google Classroom. Over 2,300 students across 69 countries, 42 of which were in Africa applied. We accepted and trained 740 students and 62\% completed. We then used the learnings to build our AI-powered smartphone app, SuaCode and then launched it in 2022 to scale the impact of SuaCode.

These courses were structured in cohorts that created a sense of community and fulfilled learners’ need for affiliation, support, and interaction and also optimized to encourage completion resulting in high completion rates (62\%, n=1000) vs industry standards (less than 20\%). Our 600+ past learners significantly improved their understanding of fundamental coding concepts despite using only smartphones to learn. We collected qualitative and quantitative feedback from our learners in various cohorts. Our quantitative analysis showed that students had an average of 17 out of 20 across all 4 assignments indicating mastery of the content \cite{boateng2021}. Also, there was no statistically significant difference in assignment scores between males and females, and educational level (high school, university, and high school graduates) suggesting that our course might be adequate for different demographics \cite{boateng2021}. Furthermore, there was a statistically significant improvement in students’ self-reported proficiency in the programming concepts \cite{boateng2021}. In user surveys about the smartphone coding experience, 85\% of our learners (n=457) rated the coding experience 4+ on a 5-point Likert scale \cite{suacodeafrica2} with feedback such as \textit{“It was really convenient, honestly. I didn’t have to necessarily sit behind a desk to do it so I could do it when I was on my bed, eating, even using the bathroom. So it was fun and convenient coding on my phone”}. Overall, there were several positive feedbacks about the experience such as \textit{“Suacode has been a very great experience for me. I got to learn processing and actually code on my phone. I also had help from the tutors and my fellow course mates which made it easier. I learnt a lot and I’m glad I had the opportunity to be part of the first batch of suacode initiative”} and \textit{“SuaCode helped improve my algorithmic thought process. I had lots of practice with thinking in a step by step process and working through challenges”}.

An innovation like SuaCode is an example of ways to get around the previously mentioned challenge of limited access to resources such as computers and infrastructure like affordable Internet which could hamper efforts to deploy AIED tools. By leveraging tools that are much more accessible to Africans like smartphones, AIED tools could to made available to these students and eventually improve their learning outcomes like we have done with SuaCode. Furthermore, our use of lesson notes rather than videos for lesson delivery addresses the challenge of expensive Internet data for African students. Intentional design of educational experiences could help to address the effect of some of the infrastructural challenges in Africa.

\subsection{AutoGrad} 
\begin{figure}[ht]
    \centering
    \includegraphics[width=\linewidth]{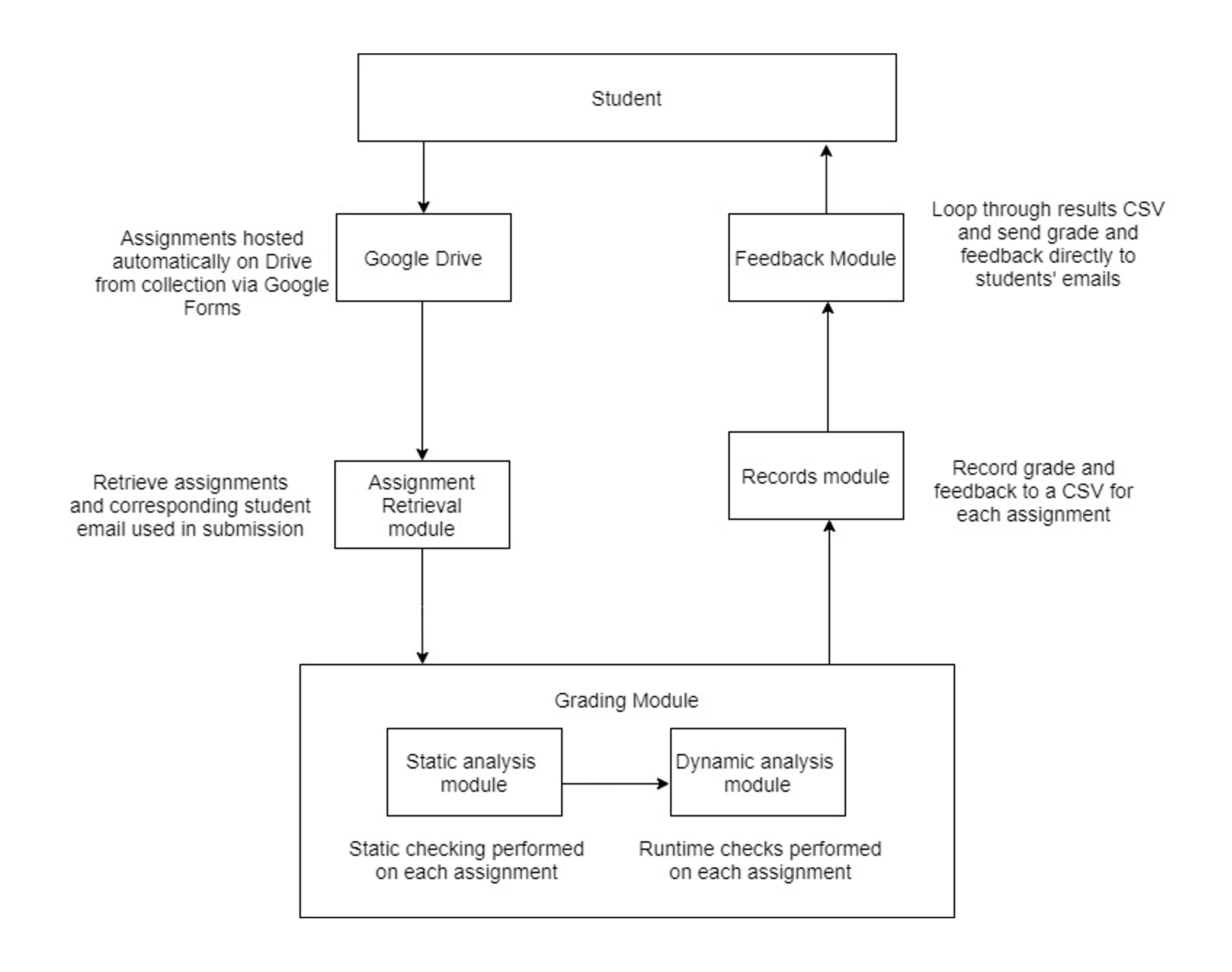}
    \caption{System design of AutoGrad (Source: \cite{annor2021})}
    \label{fig:autograd}
\end{figure}

AutoGrad is a novel cross-platform software for automatically grading and evaluating graphical and interactive programs written in the Processing programming language in SuaCode courses. \cite{annor2021}. It uses APIs to retrieve assignments from the course platform, conducts both static and dynamic analyses on these assignments to assess their graphical and interactive outputs, and subsequently furnishes students with grades and feedback (Figure \ref{fig:autograd}). The AutoGrad system itself was written using Python and Processing. The Python component manages assignment retrieval and the dissemination of grades and feedback to students, while Processing oversees the grading module, ensuring the effective execution of checks on Processing code. AutoGrad has been successfully deployed in multiple iterations of SuaCode cohorts, servicing over 1,000 students across Africa and evaluating more than 3,000 code files. These deployments involved running AutoGrad as software on a computer. Notably, in the latest cohort, students were allowed to submit complaints in cases where they believed their assignments were inaccurately graded. This approach not only ensured fairness to all students but also provided valuable insights for addressing instances where AutoGrad's performance fell short and facilitated ongoing enhancements to the software.

We assessed AutoGrad’s grading accuracy using both test assignment scripts and actual student scripts from previous cohorts of the SuaCode course. Using 10 student scripts for Assignment 1, AutoGrad yielded identical grades to those assigned by manually grading instructors in 8 scripts, resulting in a mean absolute error of 0.7. However, subsequent assignments exhibited a higher frequency of discrepancies than Assignment 1 \cite{annor2021}. We also collected feedback from students regarding their experiences with AutoGrad. Quantitative feedback was gathered by asking students to rate their agreement with the statement “I liked the feedback from AutoGrad” on a 5-point Likert scale, ranging from strongly disagree to strongly agree. Among the 457 students who completed the course and responded in the 2020 cohort, 75.9\% agreed or strongly agreed with the statement, indicating that a majority of students found AutoGrad's feedback beneficial, albeit with some room for improvement, as indicated by a mean rating of 4 out of 5. Qualitative feedback was also collected, focusing on suggestions for enhancing AutoGrad's feedback mechanism. A common theme among responses was the request for more detailed explanations accompanying the feedback provided by AutoGrad, as well as the identification of specific areas within their code where improvements could be made \cite{annor2021}.

\subsection{Code Plagiarism Detectector} 
\begin{figure}[ht]
  \centering
  \includegraphics[width=\textwidth]{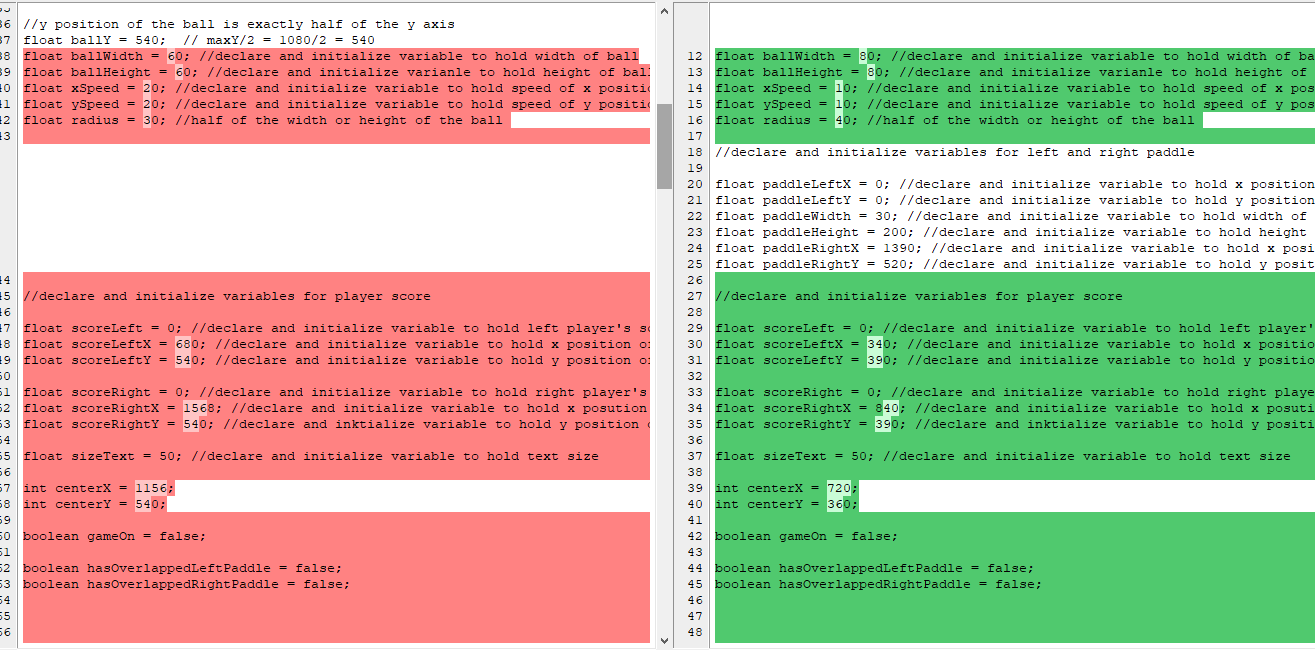}
  \caption{GUI tool highlighting plagiarized code sections in two files (Source: \cite{john2021})}
\label{fig:visual_evidence}
\end{figure}

We developed a tool that performs code plagiarism detection in students’ assignment submissions in SuaCode courses and also shows visual evidence of the detected plagiarism \cite{john2021}. We built this tool to address the issues of code plagiarism in the SuaCode where we previously identified that 27\% of the 431 students had cases of plagiarism via manual inspection. We trained machine learning models on three (3) cosine similarity-based scores extracted from the TF-IDF (with n-grams) feature vector of the code files to detect pairs of code files that had cases of plagiarism. We used as features the cosine similarity of  (1) student 1 code and student 2 code, (2) student 1 code and example code template, and (3) student 2 code and example code template. Our evaluation using 431 code files showed a balanced accuracy of 84\% using random forest with less than 1\% false positive. Also, the system provides proof of plagiarism via a GUI tool that displays side-by-side pairs of code files while highlighting sections with overlapping code. This work is the first that builds a complete end-to-end system that uses (1) machine learning algorithms to detect plagiarized source code containing English and French texts while taking the code example provided by the instructors into consideration and (2) provides visual evidence of the plagiarism (Figure \ref{fig:visual_evidence}) \cite{john2021}. 

\subsection{Kwame} 
\begin{figure}[ht]
  \centering
  \includegraphics[width=\linewidth]{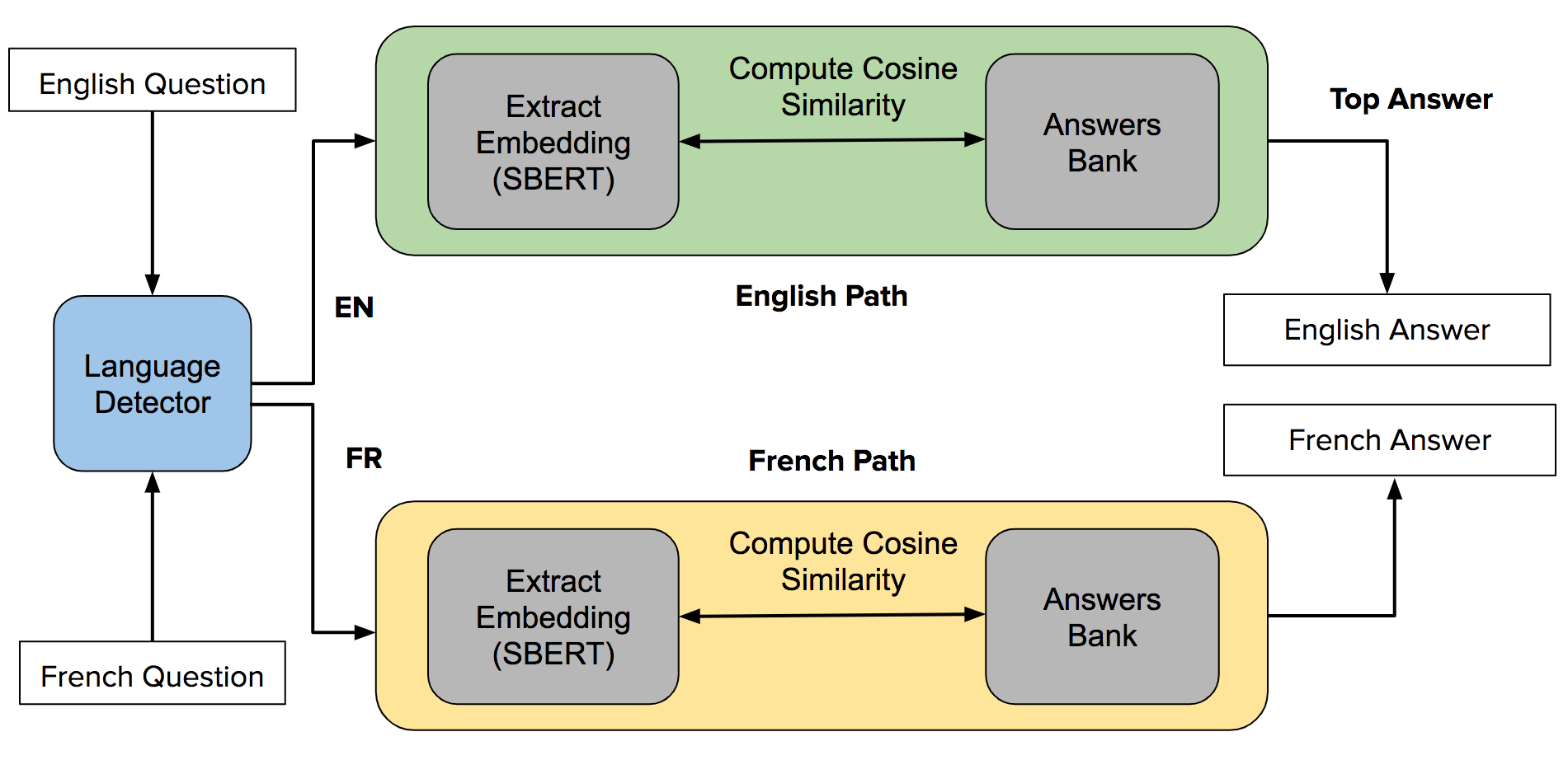}
  \caption{System Architecture of Kwame (Source: \cite{boateng2021b})}
  \label{fig:kwame_architecture}
\end{figure}

Kwame is a bilingual AI teaching assistant that provides answers to students’ coding questions in English and French for SuaCode courses \cite{boateng2021b}. Kwame is a deep learning-based question-answering system that was trained using the SuaCode course materials (lesson notes and past questions and answers) and evaluated using accuracy and time to provide answers. It finds the paragraph most semantically similar to a question via cosine similarity using Sentence-BERT (SBERT) \cite{reimers2019}, a large language model that uses siamese and triplet network architecture with BERT to train models such that semantically similar sentences are closer in vector space (Figure \ref{fig:kwame_architecture}). We compared Kwame with other approaches and performed a real-time implementation which showed fast response time and superior accuracy of top 1, 3, and 5 accuracies of 58.3\% (58.3\%), 83.3\% (80\%) and 100\% (91.7\%) for English (French) respectively. Kwame has been integrated into the SuaCode app and answers students' questions by posting an answer on the course forum in reply to students’ questions. Kwame’s pipeline is a retrieval system that consists of an ElasticSearch vector store of the embeddings of passages from the lesson notes. When a question is asked, our system computes an embedding of the question using SBERT, computes cosine similarity scores with all saved embeddings, and retrieves the top 3 passages which are then posted on the forum as answers. Kwame is named after Dr. Kwame Nkrumah the first President of Ghana and a Pan-Africanist whose vision for a developed Africa motivates this work. 

Kwame was developed to address the issue where our learners needed a lot of assistance given it was the first coding course for most learners. We relied on human facilitators to provide support and answer students’ questions. For example, in SuaCode Africa 2.0, facilitators contributed over 1,000 hours of assistance time for 8 weeks and helped to achieve an average response time of 6 minutes through the course \cite{suacodeafrica2}. This approach was however not scalable as the number of students applying to SuaCode increased exponentially year on year. Hence, in 2020, we built Kwame, an AI teaching assistant to provide accurate and quick answers to students' questions which would reduce the burden on human teaching assistants, and provide an opportunity to scale learning support. Kwame addresses the diversity of languages used for learning in Africa by offering answers in both English and French. This caters to both Anglophone and Francophone regions, covering the official languages spoken across different parts of the continent.

\subsection{Kwame for Science}
Kwame for Science \footnote{\href{http://ed.kwame.ai/}{http://ed.kwame.ai/}} is an AI-powered web application that offers two primary functionalities: (1) question answering and (2) viewing past national exam questions \cite{boateng2022, boateng2023}. These features  are enabled by our curated knowledge base of content from textbooks and past national exams over the past 28 years (for Integrated Science at the Senior High School level) with answers from certified teachers. We built Kwame for Science to address the impact of the shortage of qualified teachers \cite{UNESCO2020} across Africa which makes it difficult for students to have adequate learning support.

\begin{figure}[ht]
\includegraphics[width=\linewidth]{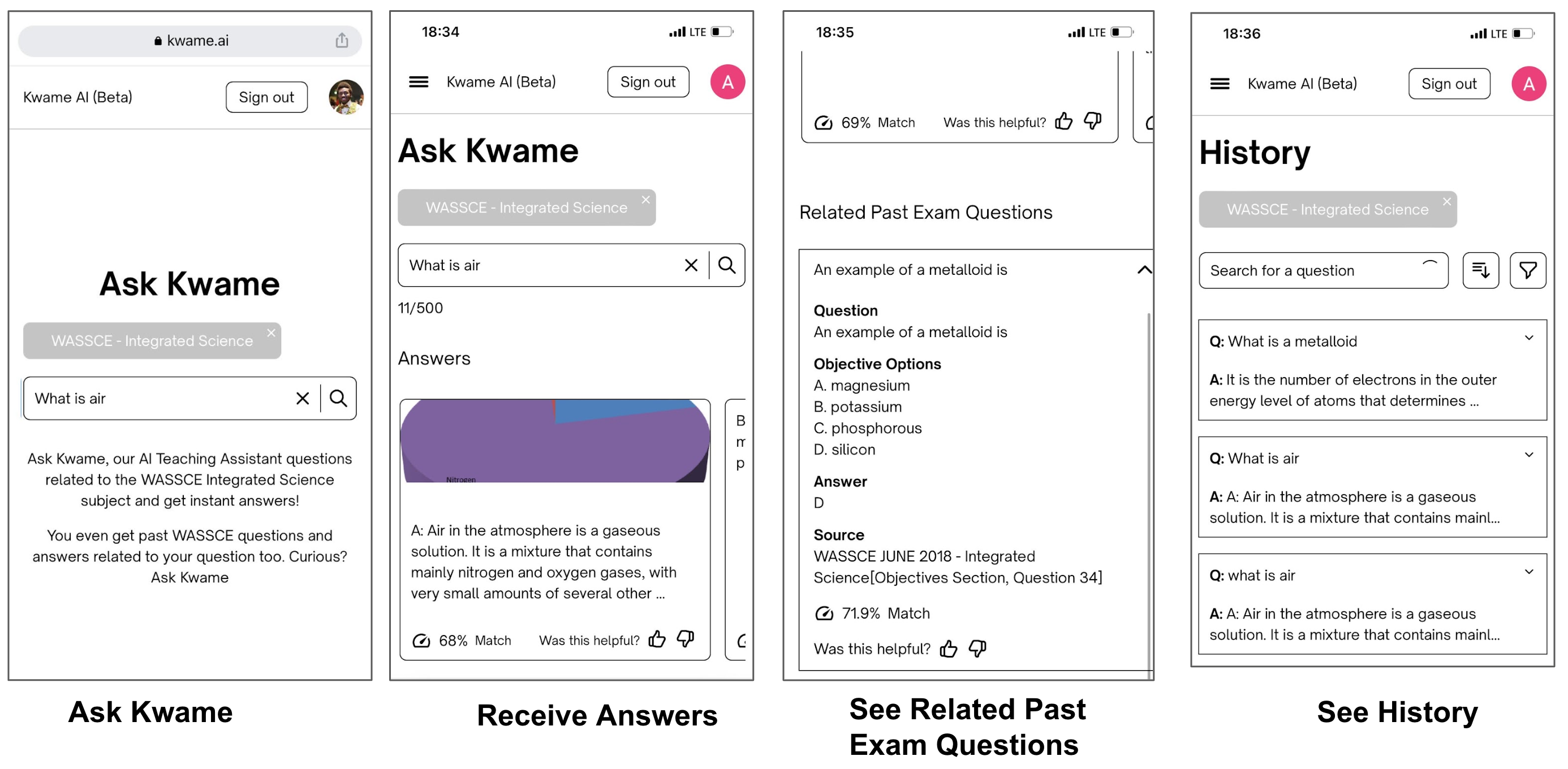}
\caption{Screenshots of QA feature of Kwame for Science (Source: \cite{boateng2023})} 
\label{fig:kwame4science}
\end{figure}

The question-answering (QA) feature allows students to pose science-related queries and receive three passages as responses, each accompanied by a confidence score (Figure \ref{fig:kwame4science}). Additionally, this feature provides the top five related past national exam questions from the Integrated Science subject of the West African Secondary School Certificate Exam (WASSCE), along with their corresponding expert answers. Users also can review their question history. When a student asks a question, our system extracts an embedding from the text using SBERT, computes cosine similarity between the embedding and the embeddings of passages from textbooks stored in ElasticSearch on Google Cloud Platform, and then returns the top answers based on the cosine similarity scores. Students could rate the helpfulness of answers and related questions.

\begin{figure}[ht]
\includegraphics[width=\linewidth]{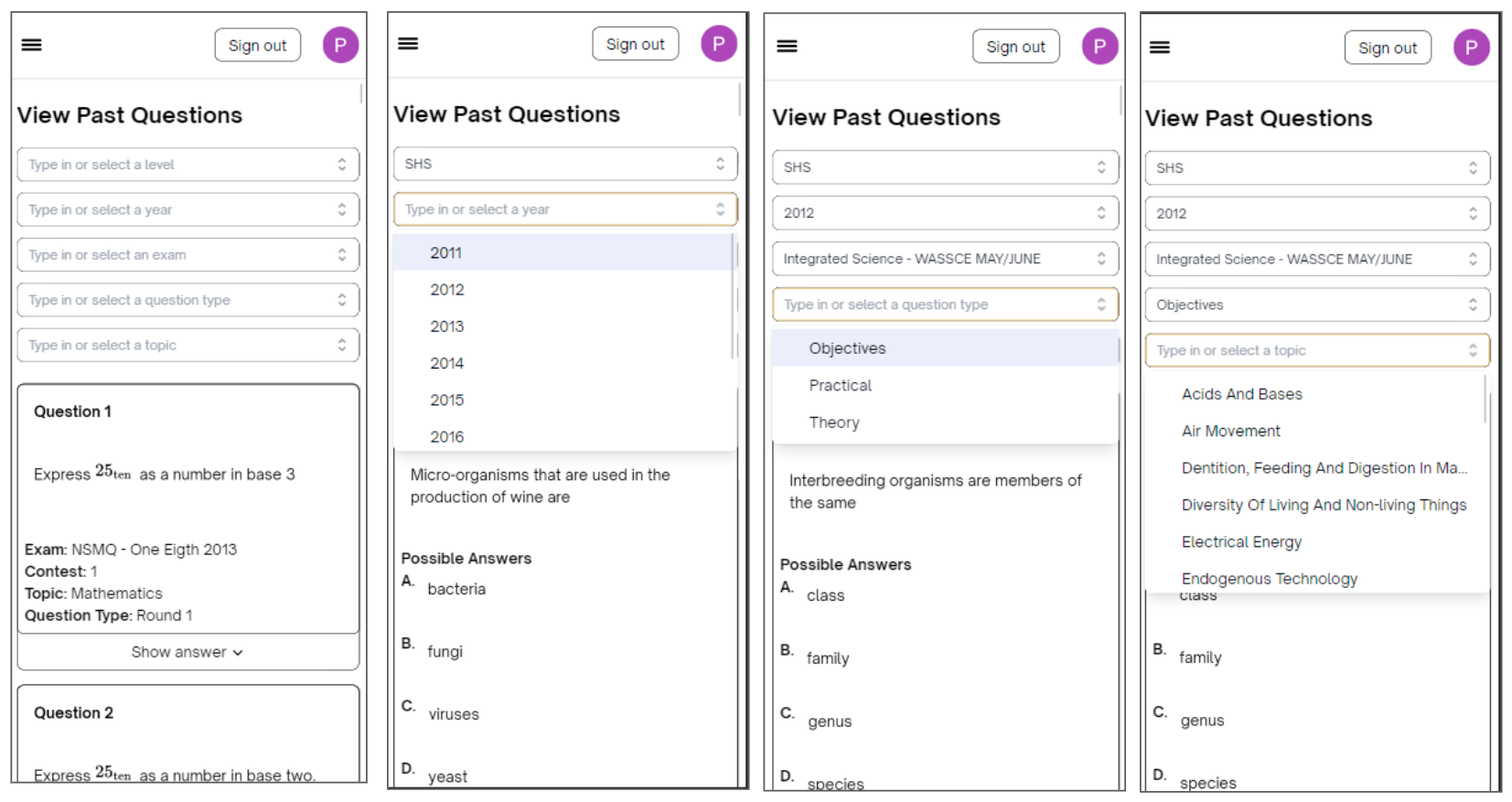}
\caption{Screenshots of View Past Questions feature of Kwame for Science (Source: \cite{boateng2023})} 
\label{fig:past_questions}
\end{figure}

Moreover, the View Past Questions feature permits students to explore past national exam questions and answers from the Integrated Science subject. This feature includes filters for refining the displayed questions based on parameters such as examination year, specific exam, question type, and automatically categorized topics generated by a custom topic detection model (Figure \ref{fig:past_questions}). All these criteria could be inferred easily from the metadata of the original exam files except the topic for which we developed a model to automatically categorize each question according to one of the syllabus topics \cite{boateng2023}. We trained a machine-learning model that used a support vector machine and SBERT embeddings of passages from the  Integrated Science subject syllabus to classify each of the past exam questions into one of the 48 syllabus topics. We then used the model to automatically categorize questions into topics in the syllabus for all 28 years of exams.  

We launched the web app in beta from 10th June 2022 to 19th February. During the 8-month deployment, we had 750 users across 32 countries (15 in Africa) asking 1.5K questions with Kwame’s helpfulness scores being top 1 and top 3 of 72.6\% and 87.2\% respectively. For the viewing past exam questions features, users most frequently used the filtering by year feature (237 times) \cite{boateng2023}. Future work will assess how Kwame for Science can improve learning outcomes.

In building Kwame for Science, we faced challenges (previously highlighted) related to accessing and using educational content from Ghana. We were unable to get official partnerships with local textbook publishers to use their copyrighted Science textbooks due to trust issues in the ecosystem. We addressed this issue by using global open-source Science textbooks and hiring local experts to provide answers to past exam questions. Furthermore, past national exam questions were only available in hardcopy formats and  open-source OCR technologies we experimented with proved inadequate in accurately extracting scientific and mathematical symbols and equations from scanned copies of the documents. We addressed this issue by hiring individuals to annotate the content and exploring commercial solutions with advanced AI technologies. Future work will explore the use of Generative AI to generate contextual, local content based on local syllabi and content from our experts, and also extract well-formatted content from scanned documents.
 
\subsection{Billa AI} 
Brilla AI is an AI contestant that we developed and deployed to unofficially compete remotely and live in the Riddles round of the 2023 NSMQ Grand Finale, the first of its kind in the 30-year history of the competition \cite{boateng2023nsmqai, boateng2024}. This work is motivated by the lack of enough qualified teachers in Africa \cite{UNESCO2020} which hampers the provision of adequate learning support. An AI could potentially augment the efforts of the limited number of teachers, leading to better learning outcomes. Yet, there exist no robust, real-world benchmark to evaluate such an AI. Towards that end, we built Brilla AI as the first key output for the \textbf{NSMQ AI Grand Challenge}, which proposed a robust, real-world challenge in education for such an AI: \textit{“Build an AI to compete live in Ghana’s National Science and Maths Quiz (NSMQ) competition and win — performing better than the best contestants in all rounds and stages of the competition”} \cite{boateng2023nsmq}. The NSMQ is an annual live science and mathematics competition for senior secondary school students in Ghana in which 3 teams of 2 students compete by answering questions across biology, chemistry, physics, and math in 5 rounds over 5 progressive stages until a winning team is crowned for that year \cite{NSMQ}. 

\begin{figure}[ht]
\includegraphics[width=\linewidth]{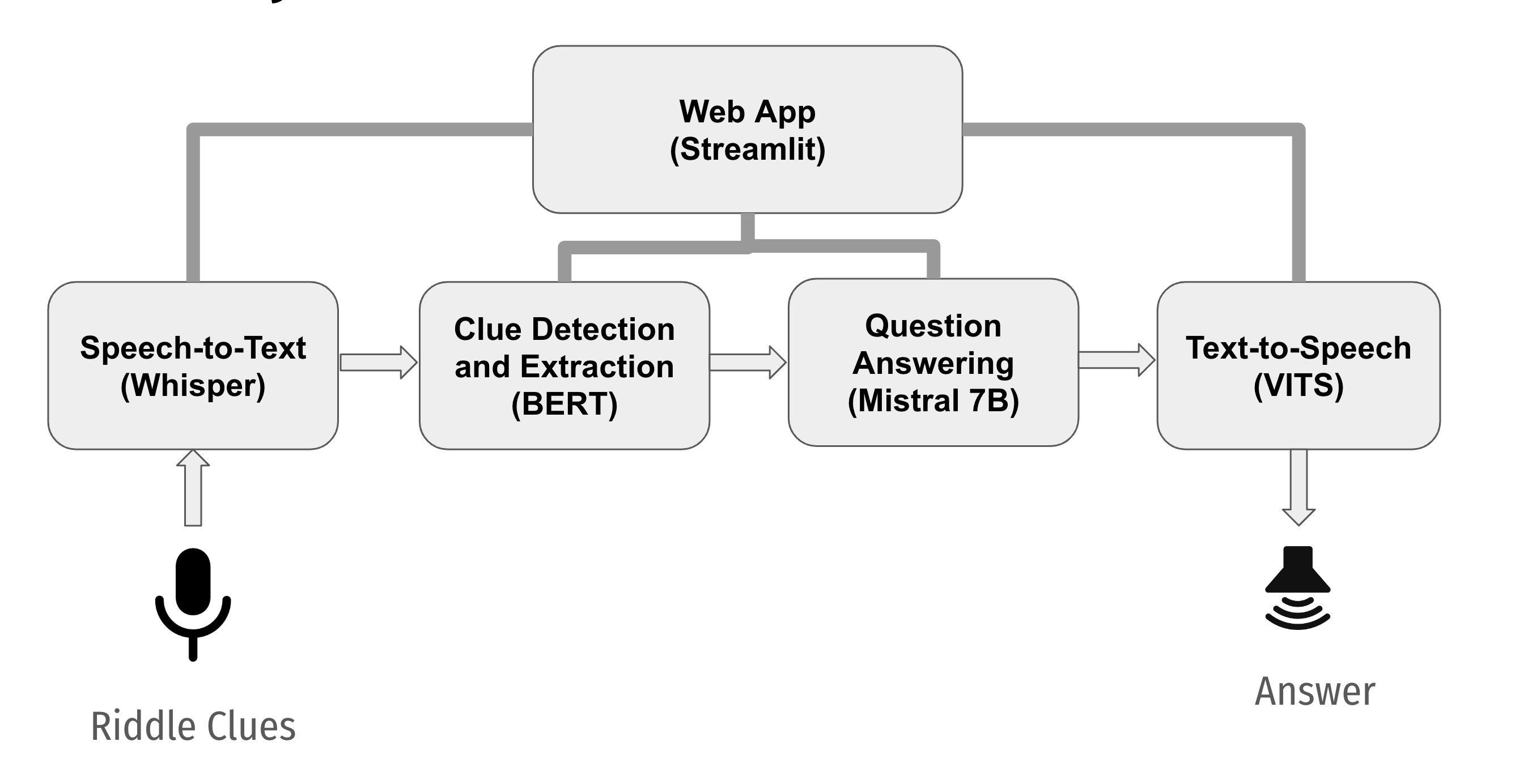}
\caption{Brilla AI System (Source: \cite{boateng2024})} 
\label{fig:brilla_ai}
\end{figure}

Brilla AI is currently available as an open-source web app (built with Streamlit \cite{streamlit}) that live streams the Riddles round of the contest, and runs 4 machine learning systems: (1) speech-to-text (using Whisper \cite{radford2023}) (2) question extraction (using BERT \cite{devlin2019}) (3) question answering (using Mistral \cite{jiang2023}) and (4) text-to-speech (using VITS \cite{kim2021}) that work together in real-time to transcribe Ghanaian accented speech, extract the question, provide an answer, and then say it with a Ghanaian accent (Figure \ref{fig:brilla_ai}) \cite{boateng2024}. In its debut in October 2023, our AI answered one of the 4 riddles ahead of the 3 human contesting teams, unofficially placing second (tied) \cite{joynews, boateng2024}. Improvements and extensions of this AI could potentially be deployed to offer science tutoring to students and eventually enable millions across Africa to have one-on-one learning interactions, democratizing science education.

Brilla AI addresses the challenge of biased AI systems that do not work well in the African context. Brilla AI contains models such as speech-to-text systems that work for African accents, text-to-speech systems that speak with African accents, and question-answering systems that provide answers using African education materials ensuring that AIED tools address the African context.

\section{Opportunities}
There is a proliferation of mobile devices, particularly, smartphones in Africa with a projection of over 600 million smartphones in Africa by the year 2025 \cite{gsma2020}. Hence, smartphones provide a unique opportunity to deliver AIED tools just as we have done with our SuaCode, a smartphone-based app for learning to code  \cite{boateng2019}. For such a context, it is important that these tools are built to have a great mobile user experience that fits different screen sizes. 

Large language Models (LLMs) and generative LLMs in particular could be leveraged to develop AI teaching assistants \cite{boateng2021, boateng2023} to augment the efforts of teachers, and AI tutors to offer personalized learning interactions with students. Leveraging approaches such as retrieval augmented generation (RAG) \cite{lewis2020} can help to address the hallucination problem of generative AI, enabling their use to generate lesson plans, lesson content, exercises, and exam questions, all grounded in local syllabi. Furthermore, RAG can be used to generate open-source textbooks using the local syllabus as the grounding, which could then be edited by local teachers to make various sections contextually relevant. These models could also be used to better convert scanned scientific and mathematical documents into well-formatted outputs. 

Various initiatives could be run to crowdsource ideas to build and develop AIED tools in Africa. We started such an initiative called AfricAIED \footnote{\url{https://www.africaied.org}}, a workshop on AI in Education in Africa. We ran the first version in 2023 which had 35 people in attendance at Google Research Ghana in Accra, Ghana, and 40 people online \cite{africaied2023}. Initiatives like this will enable the sharing of best practices and potential pitfalls toward developing AIED tools that benefit all students across Africa. Such efforts like Brilla AI are already resulting in the building of localized models \cite{boateng2023nsmqai} that could be used to develop and deploy a conversational AI tutor available via mobile devices, even non-smartphones through calling which transcribes local accented speech, provides answers to science questions using local examples, and says them out with a local accent.

\section{Conclusion} 
In this book chapter, we discussed some key challenges with leveraging AI to improve education in Africa such as limited access to computers, affordable Internet, reliable electricity, lack of regulatory support for innovation, students’ overreliance, heterogeneity in educational systems, undigitized education materials, biased AI systems and inaccuracies of generative AI. We described our work building and deploying AIED tools in the African context to advance science and computing education. In particular, we highlighted (1) SuaCode, an AI-powered app that enables Africans to learn to code using their smartphones, (2) AutoGrad, an automated grading, and feedback tool for graphical and interactive coding assignments, (3) a tool for code plagiarism detection that shows visual evidence of plagiarism, (4) Kwame, a bilingual AI teaching assistant for coding courses, (5) Kwame for Science, a web-based AI teaching assistant that provides instant answers to students’ science questions and (6) Brilla AI, an AI contestant for the NSMQ. We described opportunities to use AI to improve education in Africa such as the proliferation of smartphones, generative AI to build AI teaching assistants and tutors, and AfricAIED, a workshop on AI in Education Africa. We are bullish on the potential of AI to transform education across Africa and enable equitable, accessible, high-quality education for millions of students. It is the reason we have been leveraging resources that are more accessible like smartphones, training machine learning models with local content, and building various AI-powered education apps for the African context since 2020, long before tools like ChatGPT made AI assistants popular.

%% file: paper.bbl

\begin{thebibliography}{35}
\ifx \bisbn   \undefined \def \bisbn  #1{ISBN #1}\fi
\ifx \binits  \undefined \def \binits#1{#1}\fi
\ifx \bauthor  \undefined \def \bauthor#1{#1}\fi
\ifx \batitle  \undefined \def \batitle#1{#1}\fi
\ifx \bjtitle  \undefined \def \bjtitle#1{#1}\fi
\ifx \bvolume  \undefined \def \bvolume#1{\textbf{#1}}\fi
\ifx \byear  \undefined \def \byear#1{#1}\fi
\ifx \bissue  \undefined \def \bissue#1{#1}\fi
\ifx \bfpage  \undefined \def \bfpage#1{#1}\fi
\ifx \blpage  \undefined \def \blpage #1{#1}\fi
\ifx \burl  \undefined \def \burl#1{\textsf{#1}}\fi
\ifx \doiurl  \undefined \def \doiurl#1{\url{https://doi.org/#1}}\fi
\ifx \betal  \undefined \def \betal{\textit{et al.}}\fi
\ifx \binstitute  \undefined \def \binstitute#1{#1}\fi
\ifx \binstitutionaled  \undefined \def \binstitutionaled#1{#1}\fi
\ifx \bctitle  \undefined \def \bctitle#1{#1}\fi
\ifx \beditor  \undefined \def \beditor#1{#1}\fi
\ifx \bpublisher  \undefined \def \bpublisher#1{#1}\fi
\ifx \bbtitle  \undefined \def \bbtitle#1{#1}\fi
\ifx \bedition  \undefined \def \bedition#1{#1}\fi
\ifx \bseriesno  \undefined \def \bseriesno#1{#1}\fi
\ifx \blocation  \undefined \def \blocation#1{#1}\fi
\ifx \bsertitle  \undefined \def \bsertitle#1{#1}\fi
\ifx \bsnm \undefined \def \bsnm#1{#1}\fi
\ifx \bsuffix \undefined \def \bsuffix#1{#1}\fi
\ifx \bparticle \undefined \def \bparticle#1{#1}\fi
\ifx \barticle \undefined \def \barticle#1{#1}\fi
\bibcommenthead
\ifx \bconfdate \undefined \def \bconfdate #1{#1}\fi
\ifx \botherref \undefined \def \botherref #1{#1}\fi
\ifx \url \undefined \def \url#1{\textsf{#1}}\fi
\ifx \bchapter \undefined \def \bchapter#1{#1}\fi
\ifx \bbook \undefined \def \bbook#1{#1}\fi
\ifx \bcomment \undefined \def \bcomment#1{#1}\fi
\ifx \oauthor \undefined \def \oauthor#1{#1}\fi
\ifx \citeauthoryear \undefined \def \citeauthoryear#1{#1}\fi
\ifx \endbibitem  \undefined \def \endbibitem {}\fi
\ifx \bconflocation  \undefined \def \bconflocation#1{#1}\fi
\ifx \arxivurl  \undefined \def \arxivurl#1{\textsf{#1}}\fi
\csname PreBibitemsHook\endcsname

\bibitem[\protect\citeauthoryear{}{2020}]{UNESCO2020Digital}
\begin{botherref}
UNESCO. Startling digital divides in distance learning emerge.
\href{https://www.unesco.org/en/articles/startling-digital-divides-distance-learning-emerge}{https://www.unesco.org/en/articles/startling-digital-divides-distance-learning-emerge}
(2020)
\end{botherref}
\endbibitem

\bibitem[\protect\citeauthoryear{}{}]{iea2023}
\begin{botherref}
SDG7: Data and Projections, IEA, Paris (2023).
\href{https://www.iea.org/reports/sdg7-data-and-projections}{https://www.iea.org/reports/sdg7-data-and-projections}
\end{botherref}
\endbibitem

\bibitem[\protect\citeauthoryear{}{2020}]{dw2020}
\begin{botherref}
Why mobile internet is so expensive in Africa (2020).
\href{https://www.dw.com/en/why-mobile-internet-is-so-expensive-in-some-african-nations/a-55483976}{https://www.dw.com/en/why-mobile-internet-is-so-expensive-in-some-african-nations/a-55483976}
(2020)
\end{botherref}
\endbibitem

\bibitem[\protect\citeauthoryear{}{2020}]{UNESCO2020}
\begin{botherref}
UNESCO. Pupil-teacher ratio Sub-Saharan Africa.
\href{https://data.worldbank.org/indicator/SE.PRM.ENRL.TC.ZS?locations=ZG}{https://data.worldbank.org/indicator/SE.PRM.ENRL.TC.ZS?locations=ZG}
(2020)
\end{botherref}
\endbibitem

\bibitem[\protect\citeauthoryear{Devlin et~al.}{2019}]{devlin2019}
\begin{bchapter}
\bauthor{\bsnm{Devlin}, \binits{J.}},
\bauthor{\bsnm{Chang}, \binits{M.-W.}},
\bauthor{\bsnm{Lee}, \binits{K.}},
\bauthor{\bsnm{Toutanova}, \binits{K.}}:
\bctitle{{BERT}: Pre-training of deep bidirectional transformers for language understanding}.
In: \beditor{\bsnm{Burstein}, \binits{J.}},
\beditor{\bsnm{Doran}, \binits{C.}},
\beditor{\bsnm{Solorio}, \binits{T.}} (eds.)
\bbtitle{Proceedings of the 2019 Conference of the North {A}merican Chapter of the Association for Computational Linguistics: Human Language Technologies, Volume 1 (Long and Short Papers)},
pp. \bfpage{4171}--\blpage{4186}.
\bpublisher{Association for Computational Linguistics},
\blocation{Minneapolis, Minnesota}
(\byear{2019}).
\doiurl{10.18653/v1/N19-1423} .
\burl{https://aclanthology.org/N19-1423}
\end{bchapter}
\endbibitem

\bibitem[\protect\citeauthoryear{Achiam et~al.}{2023}]{achiam2023}
\begin{botherref}
\oauthor{\bsnm{Achiam}, \binits{J.}},
\oauthor{\bsnm{Adler}, \binits{S.}},
\oauthor{\bsnm{Agarwal}, \binits{S.}},
\oauthor{\bsnm{Ahmad}, \binits{L.}},
\oauthor{\bsnm{Akkaya}, \binits{I.}},
\oauthor{\bsnm{Aleman}, \binits{F.L.}},
\oauthor{\bsnm{Almeida}, \binits{D.}},
\oauthor{\bsnm{Altenschmidt}, \binits{J.}},
\oauthor{\bsnm{Altman}, \binits{S.}},
\oauthor{\bsnm{Anadkat}, \binits{S.}}, et al.:
Gpt-4 technical report.
arXiv preprint arXiv:2303.08774
(2023)
\end{botherref}
\endbibitem

\bibitem[\protect\citeauthoryear{}{2023}]{duolingomax}
\begin{botherref}
Introducing Duolingo Max, a learning experience powered by GPT-4.
\href{https://blog.duolingo.com/duolingo-max/}{https://blog.duolingo.com/duolingo-max/}
(2023)
\end{botherref}
\endbibitem

\bibitem[\protect\citeauthoryear{}{2023}]{q-chat}
\begin{botherref}
Introducing Q-Chat, the world’s first AI tutor built with OpenAI’s ChatGPT.
\href{https://quizlet.com/blog/meet-q-chat}{https://quizlet.com/blog/meet-q-chat}
(2023)
\end{botherref}
\endbibitem

\bibitem[\protect\citeauthoryear{}{2023}]{cheggmate}
\begin{botherref}
Chegg announces CheggMate, the new AI companion, built with GPT-4.
\href{https://www.businesswire.com/news/home/20230417005324/en/Chegg-announces-CheggMate-the-new-AI-companion-built-with-GPT-4}{https://www.businesswire.com/news/home/20230417005324/en/Chegg-announces-CheggMate-the-new-AI-companion-built-with-GPT-4}
(2023)
\end{botherref}
\endbibitem

\bibitem[\protect\citeauthoryear{}{2023}]{khanmigo}
\begin{botherref}
Harnessing GPT-4 so that all students benefit. A nonprofit approach for equal access.
\href{https://blog.khanacademy.org/harnessing-ai-so-that-all-students-benefit-a-nonprofit-approach-for-equal-access/}{https://blog.khanacademy.org/harnessing-ai-so-that-all-students-benefit-a-nonprofit-approach-for-equal-access/}
(2023)
\end{botherref}
\endbibitem

\bibitem[\protect\citeauthoryear{Boateng et~al.}{2023a}]{boateng2023}
\begin{botherref}
\oauthor{\bsnm{Boateng}, \binits{G.}},
\oauthor{\bsnm{John}, \binits{S.}},
\oauthor{\bsnm{Boateng}, \binits{S.}},
\oauthor{\bsnm{Badu}, \binits{P.}},
\oauthor{\bsnm{Agyeman-Budu}, \binits{P.}},
\oauthor{\bsnm{Kumbol}, \binits{V.}}:
Real-world deployment and evaluation of kwame for science, an ai teaching assistant for science education in west africa.
arXiv preprint arXiv:2302.10786
(2023)
\end{botherref}
\endbibitem

\bibitem[\protect\citeauthoryear{Boateng et~al.}{2023b}]{boateng2023nsmqai}
\begin{bchapter}
\bauthor{\bsnm{Boateng}, \binits{G.}},
\bauthor{\bsnm{Mensah}, \binits{J.A.}},
\bauthor{\bsnm{Yeboah}, \binits{K.T.}},
\bauthor{\bsnm{Edor}, \binits{W.}},
\bauthor{\bsnm{Mensah-Onumah}, \binits{A.K.}},
\bauthor{\bsnm{Ibrahim}, \binits{N.D.}},
\bauthor{\bsnm{Yeboah}, \binits{N.S.}}:
\bctitle{Towards an ai to win ghana’s national science and maths quiz}.
In: \bbtitle{Deep Learning Indaba 2023}
(\byear{2023})
\end{bchapter}
\endbibitem

\bibitem[\protect\citeauthoryear{}{2023}]{wasscecheating}
\begin{botherref}
Some subject results of candidates from 235 schools withheld for using AI-generated answers.
\href{https://www.myjoyonline.com/some-subject-results-of-candidates-from-235-schools-withheld-for-using-ai-generated-answers/}{https://www.myjoyonline.com/some-subject-results-of-candidates-from-235-schools-withheld-for-using-ai-generated-answers/}
(2023)
\end{botherref}
\endbibitem

\bibitem[\protect\citeauthoryear{Annor et~al.}{2021}]{annor2021}
\begin{bchapter}
\bauthor{\bsnm{Annor}, \binits{P.S.}},
\bauthor{\bsnm{Kayang}, \binits{E.}},
\bauthor{\bsnm{Boateng}, \binits{S.}},
\bauthor{\bsnm{Boateng}, \binits{G.}}:
\bctitle{Autograd: Automated grading software for mobile game assignments in suacode courses}.
In: \bbtitle{Proceedings of the 10th Computer Science Education Research Conference},
pp. \bfpage{79}--\blpage{85}
(\byear{2021})
\end{bchapter}
\endbibitem

\bibitem[\protect\citeauthoryear{John and Boateng}{2021}]{john2021}
\begin{bchapter}
\bauthor{\bsnm{John}, \binits{S.}},
\bauthor{\bsnm{Boateng}, \binits{G.}}:
\bctitle{“i didn’t copy his code”: Code plagiarism detection with visual proof}.
In: \bbtitle{International Conference on Artificial Intelligence in Education},
pp. \bfpage{208}--\blpage{212}
(\byear{2021}).
\bcomment{Springer}
\end{bchapter}
\endbibitem

\bibitem[\protect\citeauthoryear{Boateng}{2021}]{boateng2021b}
\begin{bchapter}
\bauthor{\bsnm{Boateng}, \binits{G.}}:
\bctitle{Kwame: A bilingual ai teaching assistant for online suacode courses}.
In: \bbtitle{International Conference on Artificial Intelligence in Education},
pp. \bfpage{93}--\blpage{97}
(\byear{2021}).
\bcomment{Springer}
\end{bchapter}
\endbibitem

\bibitem[\protect\citeauthoryear{SAP}{2016}]{sap2016}
\begin{botherref}
\oauthor{\bsnm{SAP}}:
Africa Code Week - Bridging the Digital Skills Gap in Africa
(2016).
\url{https://africacodeweek.org/fr/blog/https-www.linkedin.com-pulse-africa-code-week-bridging-digital-skills-gap}
\end{botherref}
\endbibitem

\bibitem[\protect\citeauthoryear{Boateng and Kumbol}{2018}]{boateng2018}
\begin{bchapter}
\bauthor{\bsnm{Boateng}, \binits{G.}},
\bauthor{\bsnm{Kumbol}, \binits{V.}}:
\bctitle{Project iswest: Promoting a culture of innovation in africa through stem}.
In: \bbtitle{2018 IEEE Integrated STEM Education Conference (ISEC)},
pp. \bfpage{104}--\blpage{111}
(\byear{2018}).
\bcomment{IEEE}
\end{bchapter}
\endbibitem

\bibitem[\protect\citeauthoryear{Boateng et~al.}{2019}]{boateng2019}
\begin{bchapter}
\bauthor{\bsnm{Boateng}, \binits{G.}},
\bauthor{\bsnm{Kumbol}, \binits{V.W.-A.}},
\bauthor{\bsnm{Annor}, \binits{P.S.}}:
\bctitle{Keep calm and code on your phone: A pilot of suacode, an online smartphone-based coding course}.
In: \bbtitle{Proceedings of the 8th Computer Science Education Research Conference},
pp. \bfpage{9}--\blpage{14}
(\byear{2019})
\end{bchapter}
\endbibitem

\bibitem[\protect\citeauthoryear{Boateng}{2020}]{boateng2020}
\begin{botherref}
\oauthor{\bsnm{Boateng}, \binits{G.}}:
Kwame: A bilingual ai teaching assistant for online suacode courses.
arXiv preprint arXiv:2010.11387
(2020)
\end{botherref}
\endbibitem

\bibitem[\protect\citeauthoryear{Boateng et~al.}{2021}]{boateng2021}
\begin{bchapter}
\bauthor{\bsnm{Boateng}, \binits{G.}},
\bauthor{\bsnm{Annor}, \binits{P.S.}},
\bauthor{\bsnm{Kumbol}, \binits{V.W.-A.}}:
\bctitle{Suacode africa: Teaching coding online to africans using smartphones}.
In: \bbtitle{Proceedings of the 10th Computer Science Education Research Conference},
pp. \bfpage{14}--\blpage{20}
(\byear{2021})
\end{bchapter}
\endbibitem

\bibitem[\protect\citeauthoryear{}{2021}]{suacodeafrica2}
\begin{botherref}
SuaCode Africa 2.0: Teaching Coding Online to Africans using Smartphones during COVID-19.
\url{https://www.c4dhi.org/news/lecture-by-boateng-suacode-africa-20210122/}
(2021)
\end{botherref}
\endbibitem

\bibitem[\protect\citeauthoryear{Reimers and Gurevych}{2019}]{reimers2019}
\begin{botherref}
\oauthor{\bsnm{Reimers}, \binits{N.}},
\oauthor{\bsnm{Gurevych}, \binits{I.}}:
Sentence-bert: Sentence embeddings using siamese bert-networks.
arXiv preprint arXiv:1908.10084
(2019)
\end{botherref}
\endbibitem

\bibitem[\protect\citeauthoryear{Boateng et~al.}{2022}]{boateng2022}
\begin{botherref}
\oauthor{\bsnm{Boateng}, \binits{G.}},
\oauthor{\bsnm{John}, \binits{S.}},
\oauthor{\bsnm{Glago}, \binits{A.}},
\oauthor{\bsnm{Boateng}, \binits{S.}},
\oauthor{\bsnm{Kumbol}, \binits{V.}}:
Kwame for science: An ai teaching assistant based on sentence-bert for science education in west africa.
iTextbooks@ AIED
(2022)
\end{botherref}
\endbibitem

\bibitem[\protect\citeauthoryear{Boateng et~al.}{2024}]{boateng2024}
\begin{botherref}
\oauthor{\bsnm{Boateng}, \binits{G.}},
\oauthor{\bsnm{Mensah}, \binits{J.A.}},
\oauthor{\bsnm{Yeboah}, \binits{K.T.}},
\oauthor{\bsnm{Edor}, \binits{W.}},
\oauthor{\bsnm{Mensah-Onumah}, \binits{A.K.}},
\oauthor{\bsnm{Ibrahim}, \binits{N.D.}},
\oauthor{\bsnm{Yeboah}, \binits{N.S.}}:
Brilla ai: Ai contestant for the national science and maths quiz.
arXiv preprint arXiv:2403.01699
(2024)
\end{botherref}
\endbibitem

\bibitem[\protect\citeauthoryear{Boateng et~al.}{2023}]{boateng2023nsmq}
\begin{botherref}
\oauthor{\bsnm{Boateng}, \binits{G.}},
\oauthor{\bsnm{Kumbol}, \binits{V.}},
\oauthor{\bsnm{Kaufmann}, \binits{E.E.}}:
Can an ai win ghana's national science and maths quiz? an ai grand challenge for education.
arXiv preprint arXiv:2301.13089
(2023)
\end{botherref}
\endbibitem

\bibitem[\protect\citeauthoryear{}{}]{NSMQ}
\begin{botherref}
National Science and Maths Quiz.
\href{https://nsmq.com.gh/}{https://nsmq.com.gh/}
\end{botherref}
\endbibitem

\bibitem[\protect\citeauthoryear{}{}]{streamlit}
\begin{botherref}
Streamlit. A faster way to build and share data apps.
\url{https://streamlit.io/}
\end{botherref}
\endbibitem

\bibitem[\protect\citeauthoryear{Radford et~al.}{2023}]{radford2023}
\begin{bchapter}
\bauthor{\bsnm{Radford}, \binits{A.}},
\bauthor{\bsnm{Kim}, \binits{J.W.}},
\bauthor{\bsnm{Xu}, \binits{T.}},
\bauthor{\bsnm{Brockman}, \binits{G.}},
\bauthor{\bsnm{McLeavey}, \binits{C.}},
\bauthor{\bsnm{Sutskever}, \binits{I.}}:
\bctitle{Robust speech recognition via large-scale weak supervision}.
In: \bbtitle{International Conference on Machine Learning},
pp. \bfpage{28492}--\blpage{28518}
(\byear{2023}).
\bcomment{PMLR}
\end{bchapter}
\endbibitem

\bibitem[\protect\citeauthoryear{Jiang et~al.}{2023}]{jiang2023}
\begin{botherref}
\oauthor{\bsnm{Jiang}, \binits{A.Q.}},
\oauthor{\bsnm{Sablayrolles}, \binits{A.}},
\oauthor{\bsnm{Mensch}, \binits{A.}},
\oauthor{\bsnm{Bamford}, \binits{C.}},
\oauthor{\bsnm{Chaplot}, \binits{D.S.}},
\oauthor{\bsnm{Casas}, \binits{D.d.l.}},
\oauthor{\bsnm{Bressand}, \binits{F.}},
\oauthor{\bsnm{Lengyel}, \binits{G.}},
\oauthor{\bsnm{Lample}, \binits{G.}},
\oauthor{\bsnm{Saulnier}, \binits{L.}}, et al.:
Mistral 7b.
arXiv preprint arXiv:2310.06825
(2023)
\end{botherref}
\endbibitem

\bibitem[\protect\citeauthoryear{Kim et~al.}{2021}]{kim2021}
\begin{bchapter}
\bauthor{\bsnm{Kim}, \binits{J.}},
\bauthor{\bsnm{Kong}, \binits{J.}},
\bauthor{\bsnm{Son}, \binits{J.}}:
\bctitle{Conditional variational autoencoder with adversarial learning for end-to-end text-to-speech}.
In: \bbtitle{International Conference on Machine Learning},
pp. \bfpage{5530}--\blpage{5540}
(\byear{2021}).
\bcomment{PMLR}
\end{bchapter}
\endbibitem

\bibitem[\protect\citeauthoryear{}{2023}]{joynews}
\begin{botherref}
NSMQ 2023: AI answered one riddle correctly ahead of contestants in grand-finale
(2023).
\url{https://www.myjoyonline.com/nsmq-2023-ai-answered-one-riddle-correctly-ahead-of-contestants-in-grand-finale/}
\end{botherref}
\endbibitem

\bibitem[\protect\citeauthoryear{}{2020}]{gsma2020}
\begin{botherref}
Mobile Economy Sub-Saharan Africa (2020).
\url{https://www.gsma.com/mobileeconomy/wp-content/uploads/2020/09/GSMA\_MobileEconomy2020\_SSA\_Infographic.pdf}
(2020)
\end{botherref}
\endbibitem

\bibitem[\protect\citeauthoryear{Lewis et~al.}{2020}]{lewis2020}
\begin{barticle}
\bauthor{\bsnm{Lewis}, \binits{P.}},
\bauthor{\bsnm{Perez}, \binits{E.}},
\bauthor{\bsnm{Piktus}, \binits{A.}},
\bauthor{\bsnm{Petroni}, \binits{F.}},
\bauthor{\bsnm{Karpukhin}, \binits{V.}},
\bauthor{\bsnm{Goyal}, \binits{N.}},
\bauthor{\bsnm{K{\"u}ttler}, \binits{H.}},
\bauthor{\bsnm{Lewis}, \binits{M.}},
\bauthor{\bsnm{Yih}, \binits{W.-t.}},
\bauthor{\bsnm{Rockt{\"a}schel}, \binits{T.}}, \betal:
\batitle{Retrieval-augmented generation for knowledge-intensive nlp tasks}.
\bjtitle{Advances in Neural Information Processing Systems}
\bvolume{33},
\bfpage{9459}--\blpage{9474}
(\byear{2020})
\end{barticle}
\endbibitem

\bibitem[\protect\citeauthoryear{}{}]{africaied2023}
\begin{botherref}
AfricAIED 2023 Recap.
\url{https://www.africaied.org/africaied-2023/recap}
\end{botherref}
\endbibitem

\end{thebibliography}
